# Understanding Information Spreading in Social Media during Hurricane Sandy: User Activity and Network Properties

Arif Mohaimin Sadri
e-mail: sadri.buet@gmail.com
Lyles School of Civil Engineering
Purdue University, 550 Stadium Mall Drive
West Lafayette, IN 47907, USA

Samiul Hasan
e-mail: samiul.hasan@ucf.edu
Department of Civil, Environmental, and Construction Engineering
University of Central Florida,12800 Pegasus Drive, Orlando, FL 32816

Satish V. Ukkusuri
Lyles School of Civil Engineering
e-mail: sukkusur@purdue.edu
Purdue University, 550 Stadium Mall Drive
West Lafayette, IN 47907, USA

Manuel Cebrian
e-mail: Manuel.Cebrian@data61.csiro.au
Data61, CSIRO
115 Batman Street, West Melbourne VIC 3003, Australia

*Corresponding author: Satish V. Ukkusuri (Lyles School of Civil Engineering, Purdue University, 550 Stadium Mall Drive, West Lafayette, IN 47907, USA; Email: sukkusur@purdue.edu)

**Abstract**

Many people use social media to seek information during disasters while lacking access to traditional information sources. In this study, we analyze Twitter data to understand information spreading activities of social media users during hurricane Sandy. We create multiple subgraphs of Twitter users based on activity levels and analyze network properties of the subgraphs. We observe that user information sharing activity follows a power-law distribution suggesting the existence of few highly active nodes in disseminating information and many other nodes being less active. We also observe close enough connected components and isolates at all levels of activity, and networks become less transitive, but more assortative for larger subgraphs. We also analyze the association between user activities and characteristics that may influence user behavior to spread information during a crisis. Users become more active in spreading information if they are centrally placed in the network, less eccentric, and have higher degrees. Our analysis provides insights on how to exploit user characteristics and network properties to spread information or limit the spreading of misinformation during a crisis event.

**Introduction**

Communities all over the world are frequently facing disasters in many forms[1]. Natural disasters alone, over the past three decades, have caused billions of dollars in property damage and killed 2.5 million people [2,3]. The National Academies Committee on Increasing National Resilience to Hazards and Disaster established hazards resilience as a national imperative at all levels (personal, local, state, and national). Disaster resilience has received more emphasis in the domains of physical infrastructure systems and operations [4]; however, resilience should also incorporate social dimensions [5]. To promote disaster resilience and minimize the adverse impacts, communities need to have sufficient preparation and information to respond to an upcoming crisis [6-9]. Early detection of influential agents in a crisis communication network can contribute towards relaying targeted, relevant, and timely information to the vulnerable communities and concerned population-groups. As a result, any possible disruption of information flow in the network can be resisted in such a crisis ahead of time.

Effective information dissemination constitutes the key to spread awareness to every individual in a community [10-12]. It requires systematic planning, collection, organization, and delivery techniques before circulating to the target audience using different media and communication means. Online social media (such as Facebook, Twitter etc.), unlike traditional ones, can serve as alternative platforms to disseminate information during disasters. Studies have acknowledged the potential and need to efficiently analyze, record and utilize the large-scale and rich information available from these online information sources [13]. Examples of such applications can be found in many empirical studies related to emergency response [14-20], crisis informatics [21-28], and many others [29-34]. Moreover, social media connectivity and activity allow researchers to analyze and predict what happens in the real world via social network amplification [35,36].

Social science studies have reported that psychological and social factors are very important in translating hazard warning information into a collective decision [37-40]. Evacuation studies have found significant correlations of local authorities, peers, local and national media, and internet with evacuation [38]. During Hurricane Sandy, for instance, social media played an important role on information sharing. Residents from New York and New Jersey were able to receive information on smartphones using social media as they had limited access to traditional sources of information (radio, television and others) [41]. In areas without power, communications via online social media continued during and after the storm based on the continuous distribution of tweets observed throughout the city. Individuals were more likely to evacuate if they relied on social media for weather-related information during Sandy [42]. Although social media data has been analyzed for many disaster studies, a key question remains open: what is the role of the underlying network structure in spreading information in social media during disasters?

The interdependence between network topology and the function of network agents has important consequences on the robustness and resilience of real networks as they respond to random failure, targeted attacks or any other external perturbations [43]. Understanding the coupled dynamics between network structure and function has manifold applications in various fields including infrastructure systems, supply chain and logistics, biology, social and financial systems, information and communication networks, and many others [44-46]. This joint association of network structure with the entities also allows the experiment of highly dynamic behavior of the network agents that exist and interact within the complex architecture. Complex networks approaches have been used in many empirical studies of real world systems, such as, disease transmission [47,48]; transmission of computer viruses [49,50]; collapse in financial systems [51], failures of power grid [52,53]; information diffusion through social networks [54], and many others.

In this study, we investigate information spreading activities and associated network properties in social media during disasters. We have analyzed active Twitter subgraphs to understand information spreading activities of Twitter users during Hurricane Sandy. Multiple Twitter subgraphs have been created based on user activities and followee list during Hurricane Sandy. We analyze different structural properties of the subgraphs following the concepts of network science. This study contributes towards a better understanding of user activities and interactions in social media platforms during a major crisis such as hurricane Sandy. The information spreading patterns will be useful for the early detection of influential network agents in such crisis. Our quadratic prediction indicates nodes being less eccentric, more central, and having larger degrees are more capable of spreading relevant information. Such nodes, because of higher reachability to many other nodes, can make meaningful contributions in crisis contagion and help disseminate early awareness in the hurricane prone regions.

## Results

### Subgraph Construction

In order to create subgraphs at any given activity level[55], we first observe the followees of an active user and identify all the active followees of that user. We construct a directed subgraph of all active users having links from user followees directed towards the active nodes. We then observe the association between the frequencies of user activity (i.e. number of tweets during the analysis period) and network properties (both global and local) by running networks models for these subgraphs. The larger the size of the subgraphs, the more nodes it includes from a lower activity level. Fig. S1 visualizes the subgraphs of different sizes and their largest connected components. Network visualization shows highly active nodes in the active subgraphs appearing both at the largest connected component and as isolates in the periphery (Fig. S1a). Within the largest connected component, highly active nodes appear at different positions (Fig. S1b-d). The ego node of the largest hub depicts its influential position in the subgraph connectivity directing our attention towards a node-level analysis of the subgraphs (Fig. S1e).

### Activity and Degree Distributions

We analyze the relationship between user activity and the corresponding degree distributions. We obtain the best fitting to the user activity and subgraph degree distributions and a value of $x_{min}$ which refers to the minimal value of $x$ at which the power law begins to become valid [56]. User activity frequency based on all relevant keywords follows a power law distribution ($\gamma = 2.71 \pm 0.005$; $p < 0.001$; $x_{min} = 39$), whereas, activity frequency (AF) based on keywords co-appeared with 'sandy' follows a truncated power law distribution ($\gamma = 2.795 \pm 0.016$; $p < 0.001$). This indicates the existence of few nodes capable of spreading information quickly while many other nodes being less active. The degree distributions of the subgraphs at different activity frequency (AF) levels follows a truncated power law ($\gamma = 3.057 \pm 0.067$; $p < 0.01$; $AF \geq 10$). Here, $\gamma$ is the slope of the distribution. This replicates the scale-free property of many real networks having fewer nodes with larger degrees and many nodes having relatively low degree. When $\gamma$ is high, the number of nodes with high degree is smaller than the number of nodes with low degree. We may thus think that a low value of $\gamma$ denotes a more equal distribution, and higher values of $\gamma$ denote more and more unfair degree distributions. However, this might not be the case and the opposite may become true i.e. a high value of $\gamma$ represents a network in which the distribution of edges is fairer. The best fit power law may only cover a portion of the distribution's tail [56]. There are domains in which the power law distribution is a superior fit to the lognormal [57]. However, difficulties in distinguishing the power law from the lognormal are common and well-described, and similar issues apply to the stretched exponential and other heavy-tailed distributions [58,59]. Our comments on the distributions fitting are based on pairwise comparison between power law, truncated power law, lognormal, and exponential distributions. See Fig. 1 for details.

## Network Analysis

Fig. 2 shows the variation of the subgraph network properties at various activity levels. It is important to note here that the larger the size of the subgraphs, the more nodes it include from a lower activity level. We observe that the number of nodes and links generated in these subgraphs (both directed and undirected) grow exponentially for larger subgraphs. A similar pattern is observed for the nodes and links that exist in the largest connected component. There exists almost equal number of connected components and isolates for all levels of activity. Network densities of the subgraphs (both directed and undirected) tend to zero for larger subgraphs, having slightly higher densities in the largest connected component in each case. This implies that the connectivity between nodes do not follow the rate at which the network grows for larger subgraphs.

Network transitivity implies the probability of any two given nodes in the graph to be connected if they are already connected to some other node. The average clustering coefficient of the undirected subgraphs range between 0.2 to 0.4 and decreases with the size of the subgraphs (see Fig. 3). The network transitivity, based on average clustering coefficient, suggests that the subgraphs become less transitive as their size grows. The increase in average degree of the nodes is indicative of more nodes that are reachable in larger subgraphs on average. The degree pearson correlation coefficient approaches 0 for larger subgraphs. This is a measure of graph assortativity in terms of node degree and a network is said to be assortative when high degree nodes are, on average, connected to other nodes with high degree and low degree nodes are, on average, connected to other nodes with low degree. Since degree assortativity measures the similarity of connections in the graph with respect to the node degree, we observe that the networks become more assortative for larger subgraphs. While the eccentricity of a node in a graph is the maximum distance (number of steps or hops) from that node to all other nodes; radius and diameter are the minimum and maximum eccentricity observed among all nodes, respectively. For a larger subgraph, we observe that the radius takes a constant value of 5, while diameter approaches 8.

Node level properties are important to understand the role and contribution of different nodes (network agents) on the information propagation at a local scale. To obtain node level properties, we first construct an active subgraph with activity frequency (AF) ≥ 10 that includes a directed graph of 157,622 nodes and 14,498,349 links. Then we run different network models to obtain node-level properties of the undirected largest connected component with 152,933 nodes and 11,375,485 links. We observe that most of these nodes had a degree close to ~25 with activity frequency around 13. While some of the nodes, having equivalent degree, were highly active; most of the nodes in this degree region remained less active. We observe fewer nodes in the higher degree zones who remain less active than some lower degree nodes (Fig. S2-S3). However, these nodes can play important role during a crisis or emergency because of their higher access to many other nodes. Similar but less smooth trend was observed with respect to average neighbor degree. This node property is related to assortativity that measures the similarity of connections in the graph with respect to the node degree. An important insight here is that we see a chunk of nodes having very high degree neighbors who remained less active (Fig. S2-S3).

The association of activity frequency with node-level clustering coefficient and eccentricity (Fig. S4) also show well-defined range. Since eccentricity of a node is the maximum distance from that node to all other nodes, we observe that most of the nodes had an eccentricity of 6, many of them remained less active while only a few of them were highly active. More importantly, for some nodes having less eccentricity, we observed their rigidity to be less active during crisis. Fig. S4 shows that many nodes, even being part of the largest connected component, did form any cluster and remained less active. These nodes are less reachable from the nodes who are more central and form clusters. Such nodes should be given due consideration for effective information dissemination during crises. Turning to the centrality measures (Fig. S5-S6), we observe that only closeness centrality shows a well-defined range (Fig. S6). Degree centrality and eigenvector centrality shows similar patterns. Betweenness centrality suggests that almost all the nodes were having centralities equal to zero in terms of their betweenness in the network. The key take away from the centrality parameters is the pattern presented by the closeness centrality which is indicative of a lot of nodes being highly central in terms of their closeness with many other nodes who remained significantly less active.

## Information Spreading Activity of Network Agents

To assess how network agents performed in terms of spreading relevant information about Sandy, we fit both linear and quadratic models of the form:

$$y \sim \theta_0 + \theta_1 x + \epsilon$$
$$y \sim \theta_0 + \theta_1 x + \theta_1 x^2 + \epsilon$$

respectively, where $y$ represents activity frequency of nodes and $x$ is a node-level network attribute. This has been done by observing node-level network properties of the largest directed subgraph AF ≥ 10 that includes 157,622 active nodes originally and 152,933 active nodes in the largest connected component. We examined the effects of degree, in-degree, out-degree, eccentricity, and closeness centrality on spreading capacity i.e. frequency of relevant tweeting activity. The quadratic model fit the data well in each case and was chosen over the linear model. Fig 4-8 shows the maximum likelihood fit of both the quadratic and linear model (dashed line), where the shaded area is the 95% confidence interval of the quadratic model, and the dots show the performance of each of the network agents. In addition to this univariate analysis, we also run a multivariate tobit regression to determine the combined effects of network variables on information spreading. We report (Table S2) the mean, standard deviation, minimum, and maximum of the variables tested in the tobit regression (data left censored at 10). We observe higher variability of closeness centrality; however, the variability of other centrality measures is insignificant while their means close to being zero.

Our analysis indicates higher spreading activity for nodes having larger in-degrees and out-degrees in a directed network (larger degrees for undirected graph). This implies that the more information (links) a given node receives from other nodes, the more active is that node during crisis. On the other hand, a node is also highly likely to be influential in case of having more out-degrees i.e. links directing to other nodes that allow them to disseminate crisis information. The coefficient estimated for closeness centrality suggests more influential capability of a node by being more central in the active subgraph. Such nodes occupy a very convenient position in a network to be able to contribute highly in the information spreading dynamics. We also observed that less eccentric nodes are more capable of spreading information because of their higher reachability to any given node in the network. All the network variables tested under tobit regression are significant at $p<0.001$.

## Discussion

The primary focus of this study is to understand the interdependence between network topology and activities of network agents during disasters. Social communication networks play a critical role during emergencies since people may obtain weather information from traditional media such as radio or television and social media such as Facebook, Twitter, or the internet. In this study, Twitter subgraphs have been analyzed based on user activity during Hurricane Sandy to reveal the information spreading activity of network agents and the associated network properties that evolved during this major disaster. For user activity at any given level, subgraphs of social networks were constructed from the user followee list obtained at the time of data collection. For relevance, user activity was assessed on the basis of number of tweets in the data that included the word 'sandy', co-appeared with other words. For the analysis of directed graphs, we considered a network link received by user from his followee. Based on our subgraph analysis at different activity levels, we reveal several information spreading characteristics of Hurricane Sandy.

We observe that information spreading activity of nodes follows a power-law showing the existence of few nodes highly active disseminating information and many other nodes being less active. The degree distributions of the communication network also follow a power-law, executing the scale-free property of many real networks (fewer nodes with larger degrees and many other nodes with fewer degrees). Network visualization shows highly active nodes in the active subgraphs appearing both at the largest connected component and as isolates in the periphery. Within the largest connected component, highly active nodes appear at different positions. The ego node of the largest hub depicts its influential position in the subgraph connectivity directing our attention towards node level analysis.

Network analysis at different activity levels suggests that the number of nodes and links in these subgraphs (both directed and undirected) grow log-linearly with the size of subgraphs that includes close enough connected components and isolates. In contrast, the overall network connectivity (i.e. subgraph densities) tends to become zero for larger subgraphs implying that having a large number of active nodes does not help much in spreading the information or awareness even though they heavily load the network. Also, the existence of significant number of network isolates at all levels does not help in crisis because of their individual activity not contributing enough to the hazard warning dissemination. For larger subgraphs, networks become less transitive, but more assortative. This implies that active network agents are more likely to connect with similar agents (for example, having similar degrees) without contributing much in forming clusters in the neighborhood at large. The radius of the largest connect components in the larger subgraphs becomes stable at 5 that is indicative of the reachability from a given node to any other node in five steps at the maximum.

Node-level information spreading activity of network agents was assessed by running univariate linear and quadratic models with tweeting activity as a function of several topological attributes. We examined the effects of degree, in-degree, out-degree, eccentricity, and closeness centrality on spreading capacity i.e. frequency of relevant tweeting activity. The quadratic model fit the data well in each case and was chosen over the linear model. Our analysis is indicative of higher spreading capacity for nodes having larger in-degrees and out-degrees in a directed network (larger degrees for undirected graph). A node is also highly likely to be influential in case of having more out-degrees i.e. links directing to other nodes that allow them to disseminate crisis information. Nodes are more capable of spreading information if occupying more central positions in the network and being less eccentric.

This study contributes towards a better understanding of user interactions in social media platforms during a major crisis such as hurricane Sandy. The information spreading patterns will be useful for the early detection of influential network agents (more central, higher degrees, and less eccentric) in such crisis. From warning to evacuation to the post-storm recovery, such influential nodes can help disseminate more targeted information to reach out vulnerable communities at all phases of the disaster. For example, celebrities or political leaders typically occupy such positions on social media and may contribute to faster dissemination of relevant crisis information. The findings of this study are specific to hurricane Sandy, future studies should validate these results with other major hurricanes and check if such insights can be generalized to other forms of disaster. Future studies should also consider the dynamics of information spreading activities of network agents.

## Data and Methods

In 2012, residents in the coastal areas of New York and New Jersey experienced a massive storm surge produced by Hurricane Sandy, a late season hurricane causing about $50 billion in property damage, 72 fatalities in the mid-Atlantic and northeastern United States, and at least 147 direct deaths across the Atlantic basin [60]. Sandy's wind and flood are the key contributors of the heightened number of fatalities [61]. In addition, 570K buildings were destroyed, 20K flights were cancelled, and 8.6M power outages in 17 states among other direct impacts of Sandy [62]. Moreover, thousands of people were displaced from their homes [63] and 230K cars were destroyed by the floods even though the residents were given early warnings about the oncoming storm and the likely impact [64]. The specific date, time, location and event, attributed to Hurricane Sandy, are presented in Table S1. Please see [65] for further details.

Twitter users can share short messages up to 140 characters and follow other users creating a network of a large number of accounts with characteristics both of a social network and an informational network [66]. The social network properties of Twitter provides access to geographically and personally relevant information and the information network properties instigate information contagion globally [36]. These specific features make Twitter particularly useful for effective information dissemination during crises. From an emergency research perspective, many researchers used Twitter to study the service characteristics [16,24], retweeting activity [67,68], situational awareness [69,70], online communication of emergency responders [71,72], text classification and event detection [21,22,26,73,74], devise sensor techniques for early awareness [65], quantifying human mobility [75,76], and disaster relief efforts [77].

In this study, we analyze raw data (~52 M tweets, ~13 M users, Oct 14 -Nov 12, 2012) obtained from Twitter. Please see [65] for the detailed steps involved in data collection. The data includes a text database with user and text identifiers, texts, and some additional useful information. The network database includes the relationship graphs of active users i.e. the list of followees for each user. These were reconstructed using Twitter API. Only a minor fraction of the texts (~ 1.35%) are geo-tagged by Twitter. For relevance, user activity was assessed on the basis of number of tweets (~11.83 M) in the data that included the word 'sandy', co-appeared with other words after filtering out ~46.45 M tweets that are in English i.e. non-English tweets were removed. For the analysis of directed graphs, we considered a network link received by user from his followee. From Twitter perspective, a followee is the user who is being followed by another user and the information flows from the followee to the followers. In the network data, we observed that a number of highly active nodes did not have any followee, however they appeared in the followee list less active users which is indicative of the direction and rate of information flow. Some active users did not appear in the network database for which we assumed zero followee since the current length of followee list on Twitter is close to zero, even after three years of data collection.

## Additional Information

### Research Ethics
The study did not require completing an ethical assessment prior to conducting the research.

### Animal Ethics
The study did not require completing an ethical assessment prior to conducting the research.

### Permission to Carry out Fieldwork
The study did not require obtaining any permission prior to conducting the research.

### Data Availability
Data is available at Dryad: http://dx.doi.org/10.5061/dryad.15fv2

### Author Contributions Statement

All the authors have contributed to the design of the study, conduct of the research, and writing the manuscript. All authors gave final approval for publication.

### Competing Financial Interests
Authors declare no competing financial interests.

## Funding

The authors are grateful to National Science Foundation for the grant CMMI-1131503 and CMMI-1520338 to support the research presented in this paper. However, the authors are solely responsible for the findings presented in this study.

**List of Figures**

**List of Tables**

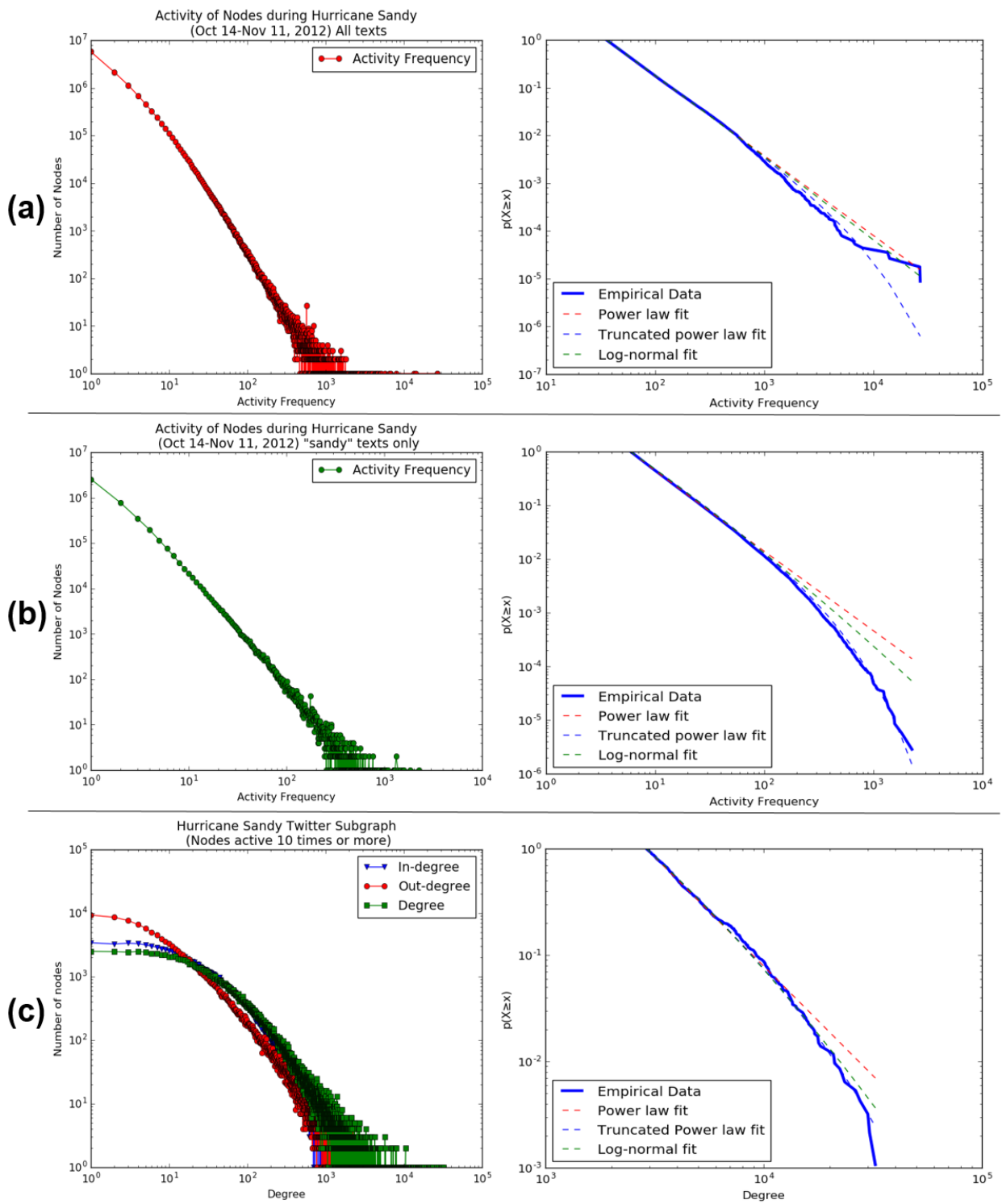

Figure 1 Activity and Degree Distributions
(a) Activity distribution based on all tweets after initial filtering (~ 46.45 M tweets). This follows a power law distribution. (b) Activity distribution of different users after the 'sandy' filtering (~ 11.83 M tweets). This follows a truncated power law distribution. (c) Degree distribution of the largest directed subgraph (~ 0.16 M nodes, ~ 14.50 M links, AF ≥ 10, ~ 3.92 M tweets)

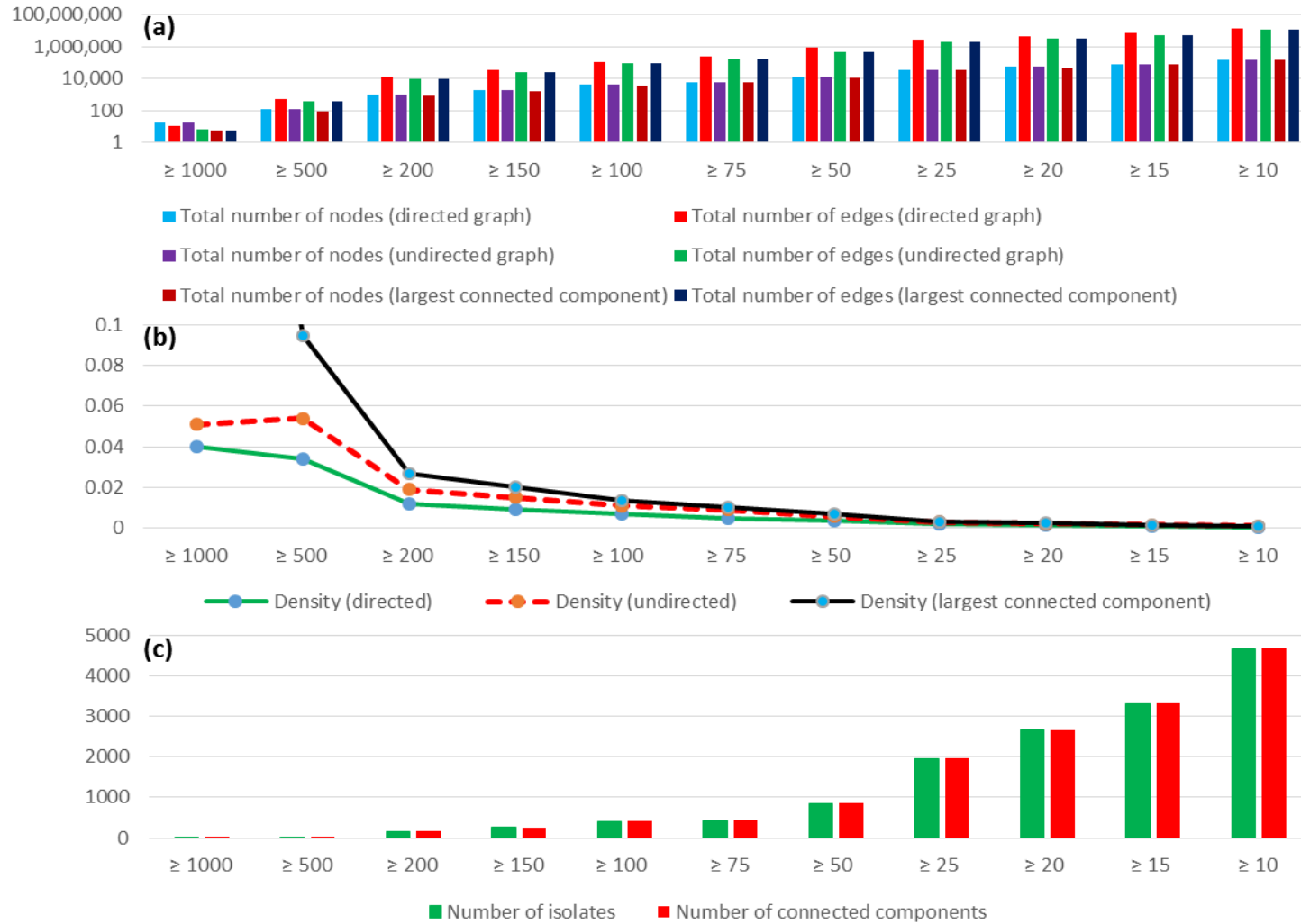

Figure 2 Subgraph properties at different activity levels (where activity level is defined by the number of tweets made by a node) (a) Number of nodes and links, (b) Network densities, (c) Number of isolates and connected components,

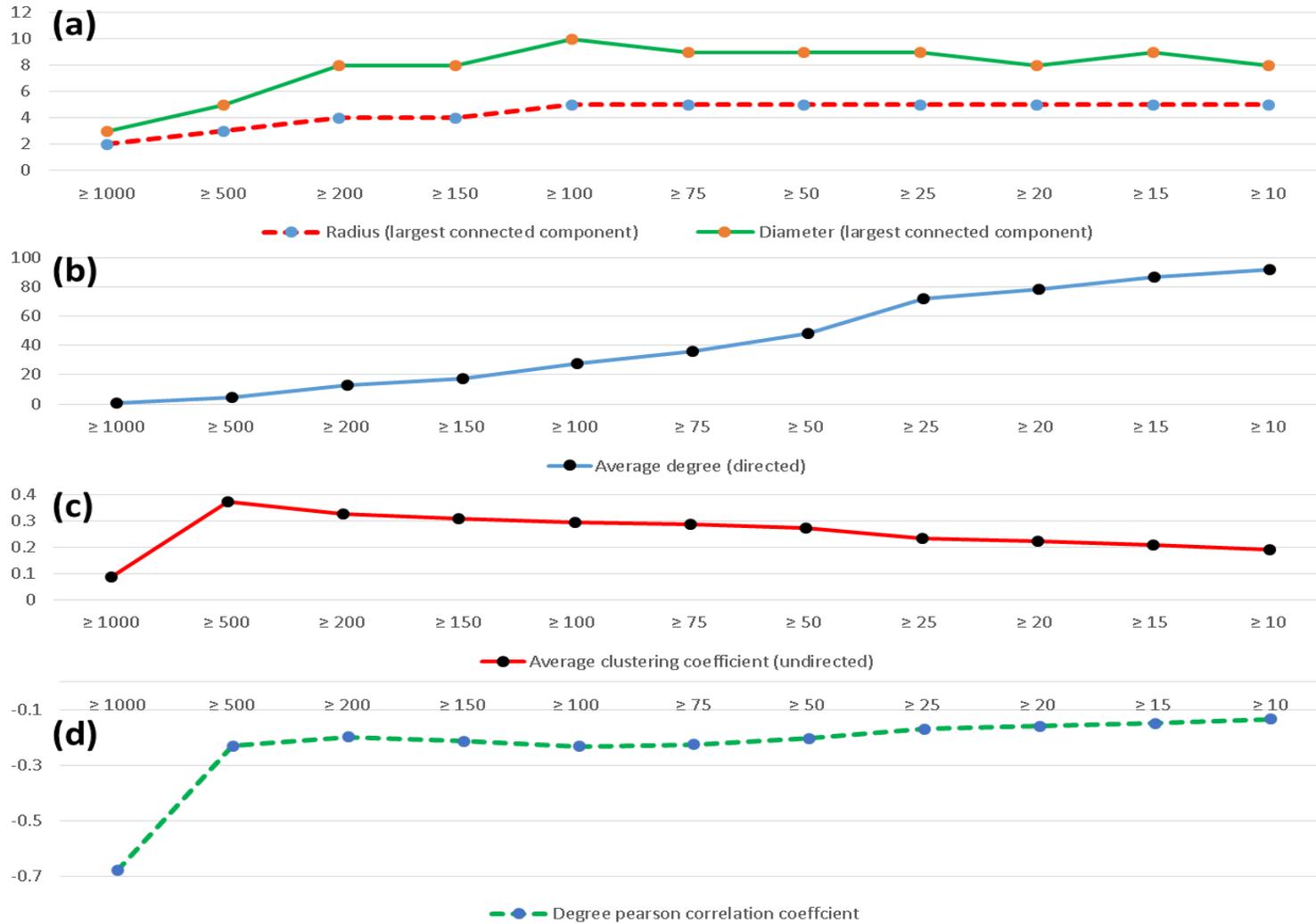

Figure 3 Subgraph properties at different activity levels

(a) Radius and Diameter of the largest connected component, (b) Average degree, (c) Average clustering coefficient, and (d) the degree

Pearson correlation coefficient

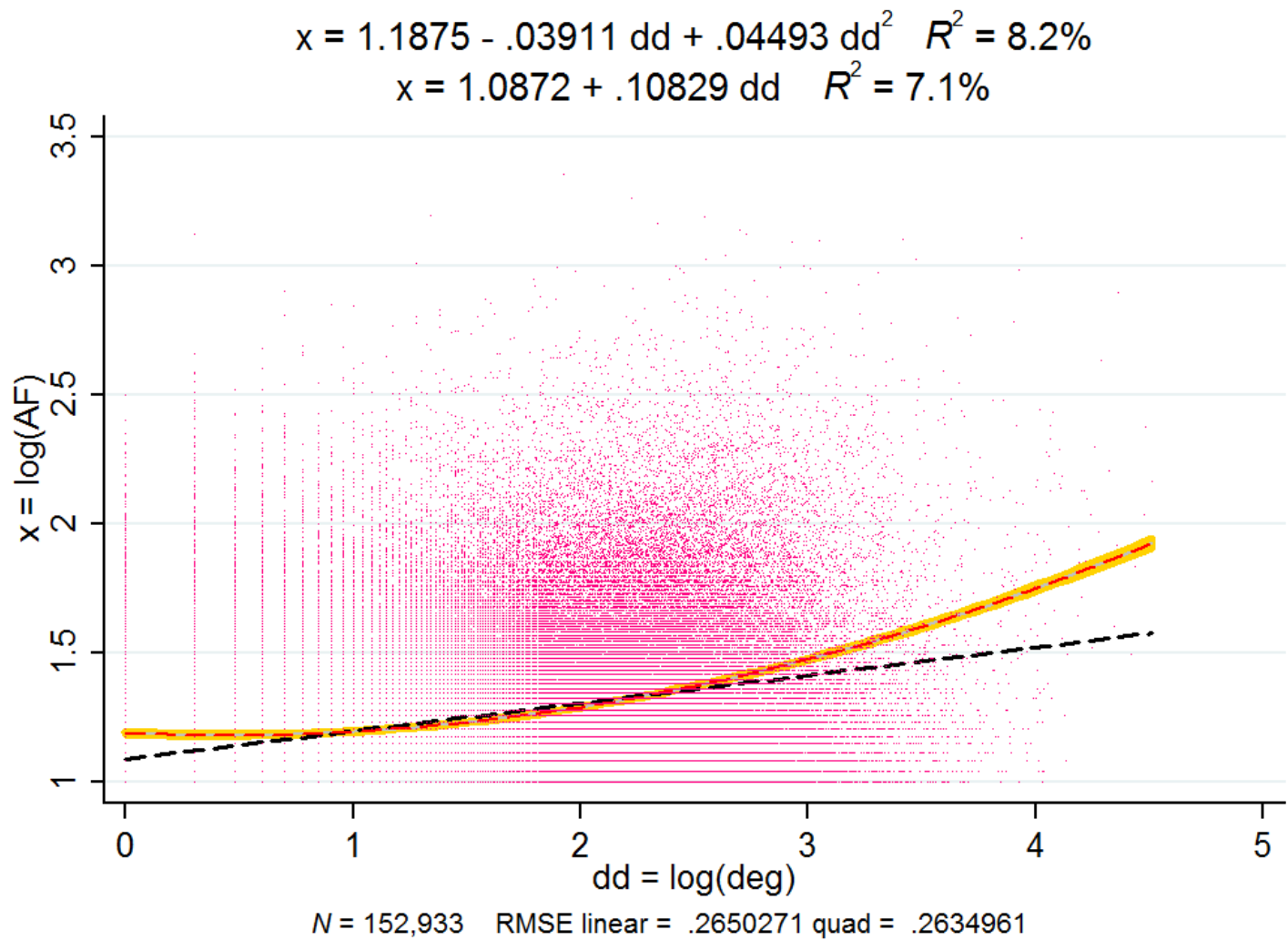

Figure 4 Information spreading capacity as a function of degree

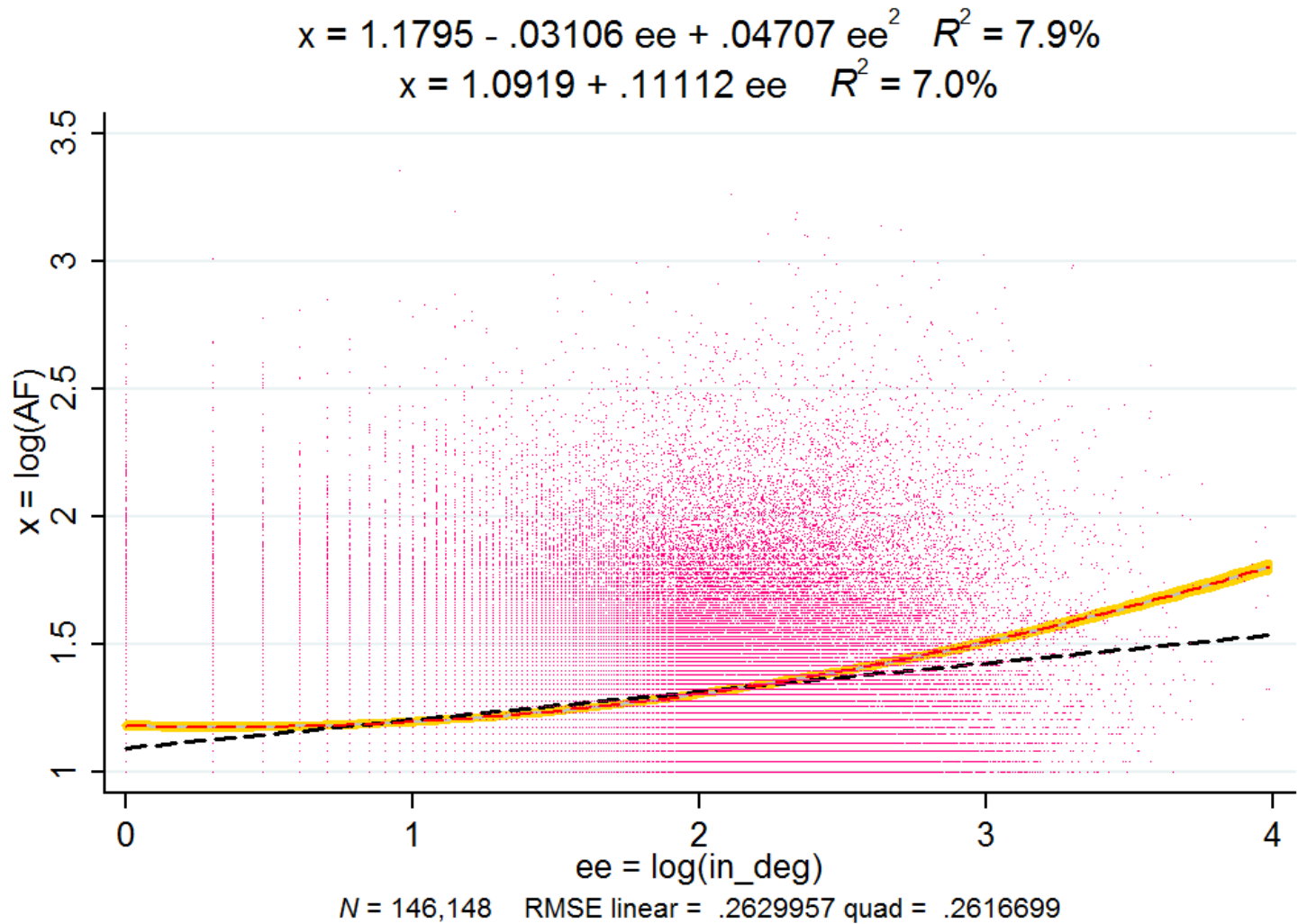

Figure 5 Information spreading capacity as a function of in-degree

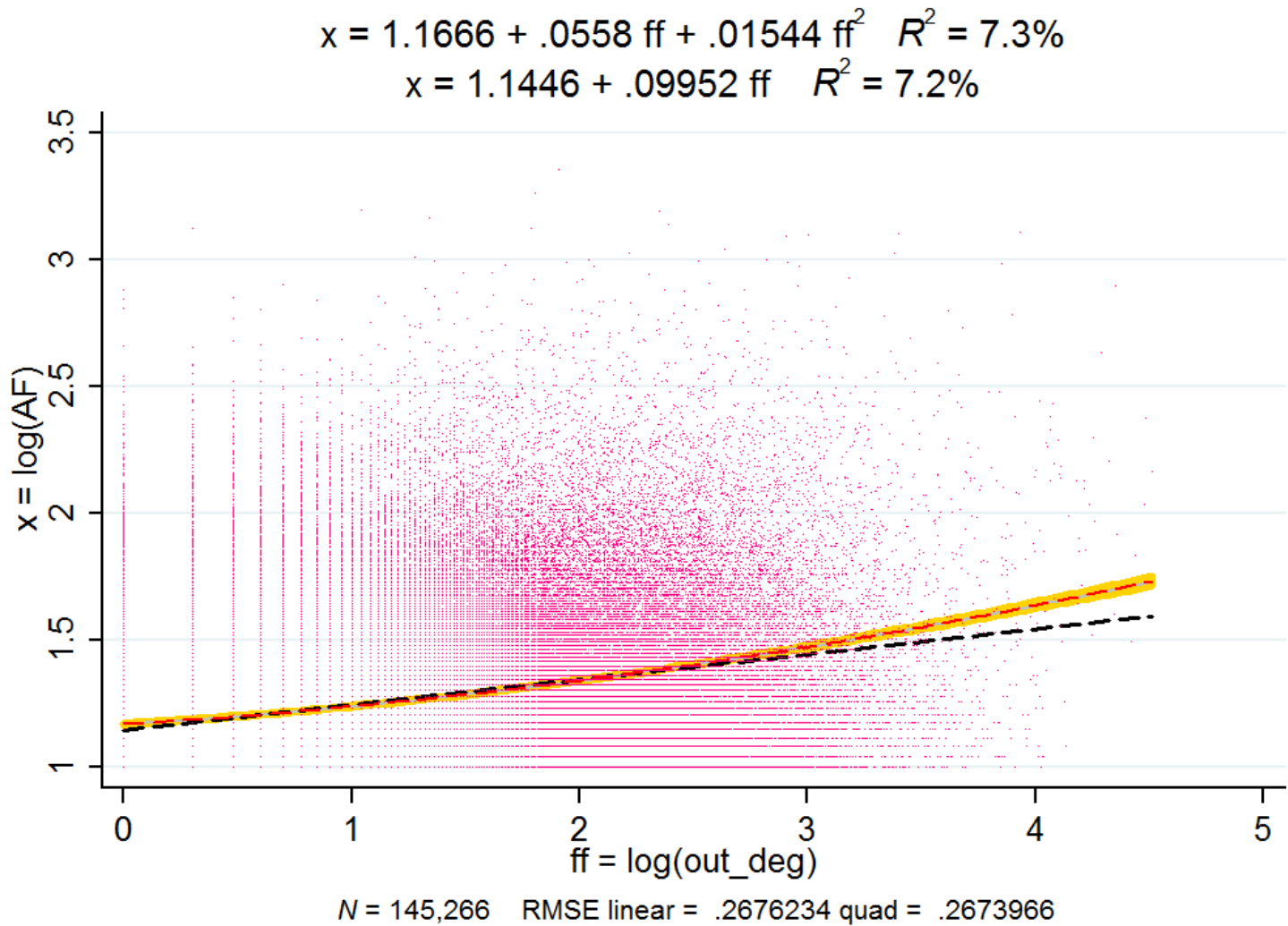

Figure 6 Information spreading capacity as a function of out-degree

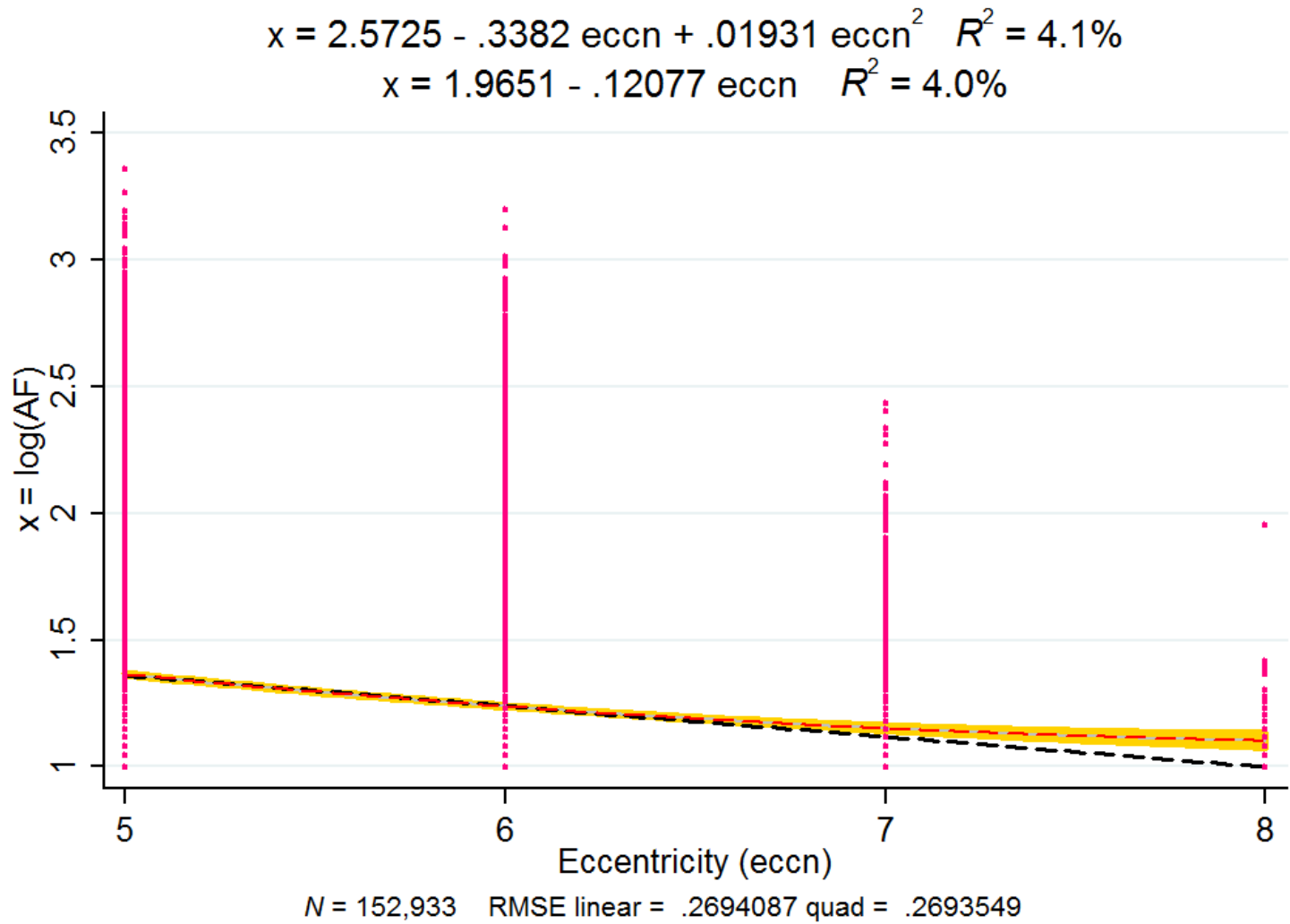

Figure 7 Information spreading capacity as a function of eccentricity

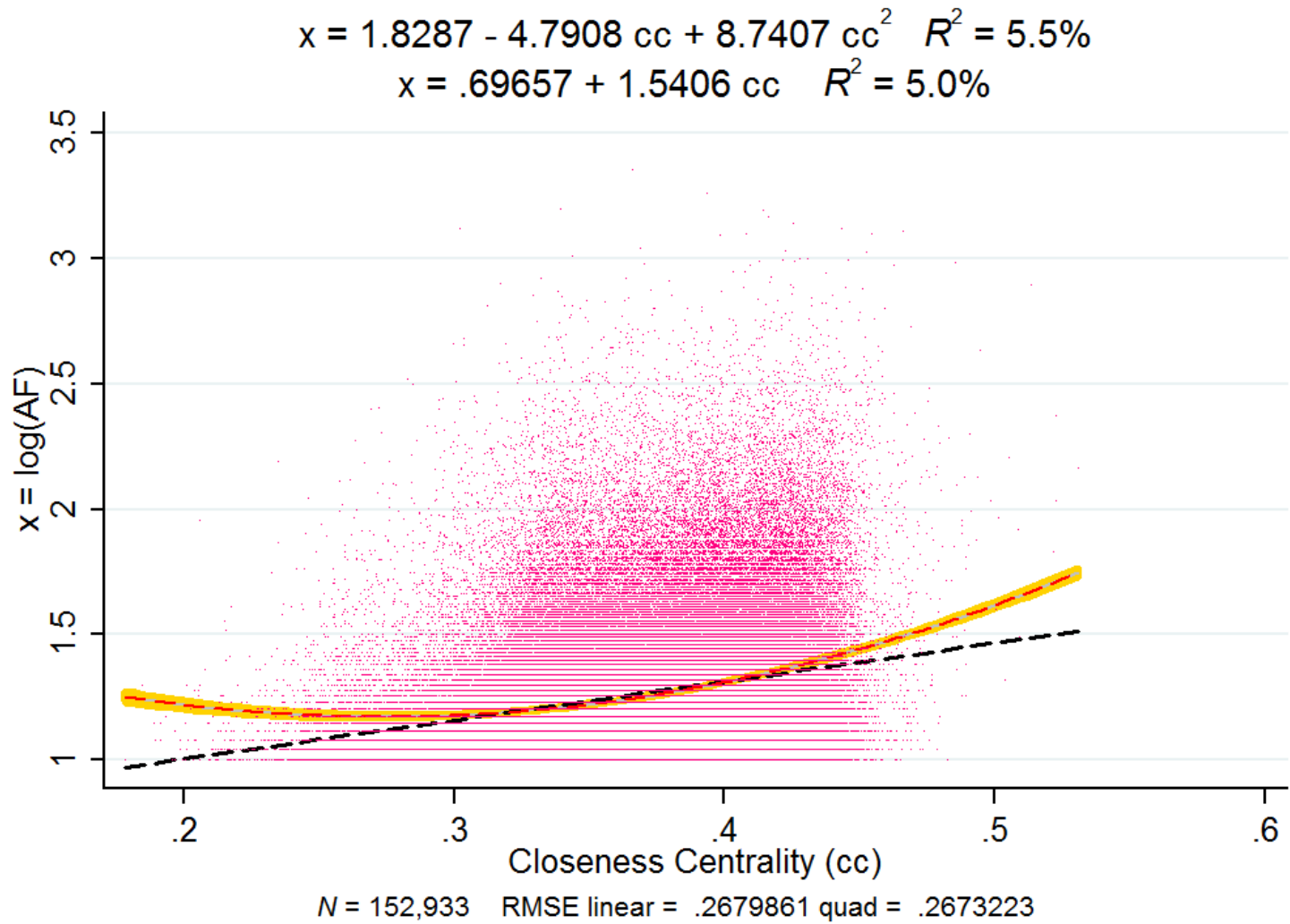

Figure 8 Information spreading capacity as a function of closeness centrality

**Table 1 Tobit Regression for Activity Frequency**

|  | *N* | *Mean* | *S.D.* | *Min.* | *Max.* |
|---|---|---|---|---|---|
| Activity Frequency | 152,933 | 24.87351 | 37.59742 | 10 | 2267 |
| Degree | 152,933 | 148.7643 | 484.8861 | 1 | 32406 |
| In-degree | 152,933 | 94.80183 | 219.0143 | 0 | 9640 |
| Out-degree | 152,933 | 94.80183 | 447.848 | 0 | 32397 |
| Clustering Coefficient | 152,933 | 0.195215 | 0.13114 | 0 | 1 |
| Eccentricity | 152,933 | 5.790215 | 0.456218 | 5 | 8 |
| Avg. Neighbor Degree | 152,933 | 2745.791 | 2326.095 | 1.5 | 32406 |
| Betweenness Centrality | 152,933 | 1.14E-05 | 0.000231 | 0 | 0.037856 |
| Closeness Centrality | 152,933 | 0.369475 | 0.040014 | 0.178652 | 0.530577 |
| Eigenvector Centrality | 152,933 | 0.00109 | 0.002313 | 8.86E-13 | 0.071367 |
| Degree Centrality | 152,933 | 0.000973 | 0.003171 | 6.54E-06 | 0.211898 |

| | **Activity Frequency: Tobit Regression** | | | | |
|---|---|---|---|---|---|
| | *Coeff.* | *Std. Err.* | *t-stat* | **[95% Conf. Int.]** | |
| Constant | 18.92990 | 2.73118 | 6.93 | 13.57685 | 24.28295 |
| In-degree | 0.00917 | 0.00057 | 16.10 | 0.008053 | 0.010285 |
| Out-degree | 0.00658 | 0.00026 | 25.63 | 0.006081 | 0.007088 |
| Eccentricity | -5.53433 | 0.30563 | -18.11 | -6.13336 | -4.9353 |
| Closeness Centrality | 87.82817 | 3.54151 | 24.80 | 80.88689 | 94.76944 |
| No. of observations | 152,933 | | | | |
| Pseudo R-squared | 0.00420 | | | | |

# Supplementary Information

## Understanding Information Spreading in Social Media during Hurricane Sandy: User Activity and Network Properties

Arif Mohaimin Sadri[1], Samiul Hasan[2], Satish V. Ukkusuri[1,*], Manuel Cebrian[3]

[1] Lyles School of Civil Engineering, Purdue University, 550 Stadium Mall Drive, West Lafayette, IN 47907, USA

[2] Department of Civil, Environmental, and Construction Engineering, University of Central Florida, 12800 Pegasus Drive, Orlando, FL 32816.

[3] Data61, CSIRO, 115 Batman Street, West Melbourne VIC 3003, Australia

[*]**Corresponding author:** Satish V. Ukkusuri (Lyles School of Civil Engineering, Purdue University, 550 Stadium Mall Drive, West Lafayette, IN 47907, USA; Email: sukkusur@purdue.edu)

## Review of Network Science

Many new network concepts, properties and measures have been developed by running experiments on large-scale real networks. A number of statistical properties and unifying principles of real networks have been identified from these studies. Significant amount of research efforts have helped to develop new network modeling tools, reproduce the structural properties observed from empirical network data, and design such networks efficiently with a view to obtaining more advanced knowledge of the evolutionary mechanisms of network growth [1]. Many real networks possess interesting properties unlike random graphs indicative of possible mechanisms guiding network formation and ways to exploit network structure with specific objectives [2]. Some of these properties, common across many real networks, are described below:

*Small-world property*: This property refers to the existence of relatively short paths between any pair of nodes in most networks despite their large size. The existence of this property is evident in many real networks [3-5]. The small-world effect has important implications in explaining dynamics of processes occurring on real networks. In case of spreading information or ideas through a network, the small-world property suggests that the propagation will be faster on most real world networks because of short average path lengths [2]. Three important measures to explain this property are eccentricity, radius and diameter. While the eccentricity of a node in a graph is the maximum distance (number of steps or hops) from that node to all other nodes; radius and diameter are the minimum and maximum eccentricity observed among all nodes, respectively.

*Degree distributions*: The degree of a node $(k)$ is the number of direct links to other nodes in a graph. The degree distribution $P(k)$ in real networks (probability that a randomly chosen node has degree $k$, issignificantly different from the Poisson distribution, typically assumed in the modeling of random graphs. In fact, real networks exhibit a power law (or scale-free) degree distribution characterized by higher densities of triangles (cliques in a social network, for example) [6]. In addition, many real networks also exhibit significant correlations in terms of node degrees or attributes. This scale-free property validates the existence of hubs, or a few nodes that are highly connected to other nodes in the network. The presence of large hubs results in a degree distribution with long tail (highly right-skewed), indicating the presence of nodes with a much higher degree than most other nodes. For an undirected network, the degree distribution $P_{degree}(k)$ can be written as follows:

$$P_{degree}(k) \propto k^{-\gamma} \dots\dots\dots\dots\dots\dots\dots\dots\dots. (2)$$

where $\gamma$ is some exponent and $P_{degree}(k)$ decays slowly as the degree $k$ increases, increasing the probability of obtaining a node with a very high degree. Networks with power-law distributions are called scale-free networks [7] that holds the same functional form (power laws) at all scales. The power law $P_{degree}(k)$ remains unchanged (other than a multiplicative factor) when rescaling the independent variable $k$ by satisfying:

$$P_{degree}(xk) = x^{-\gamma} P_{degree}(k) \dots\dots\dots\dots\dots\dots\dots. (3)$$

The presence of hubs that are orders of magnitude larger in degree than most other nodes is a characteristic of power law networks. In this study, we test the scale free property both for the activity frequency of all active nodes and the degree distribution of subgraphs being active at different activity levels.

*Transitivity*: This property is a distinctive deviation from the properties of random graphs. Network transitivity implies that two nodes are highly likely to be connected in a network, given each of the nodes are connected to some other node. This is indicative of heightened number of triangles that exist in real networks (sets of three nodes each of which is connected to each of the others) [2]. The existence of triangles can be quantified by *Clustering Coefficient. C:*

$$C = \frac{3 * \text{Number of triangles in the network}}{\text{Number of connected triples of nodes}} \quad \ldots\ldots\ldots\ldots\ldots (4)$$

A *connected triple* refers to a single node with links running to an unordered pair of others. In case of social networks, transitivity refers to the fact that the friend of one's friend is likely also to be the friend of that person. Another important notion is *Network Density*, frequently used in the sociological literature [8]. The density is 0 for a graph without any link between nodes and 1 for a completely connected graph.

*Network resilience*: This property, related to degree distributions, refers to the resilience of networks as a result of removing random nodes in the network and the level of resilience to such vertex removal varies across networks depending on the network topology [2]. Networks in which most of the nodes have low degree have less disruption since these nodes lie on few paths between others; whereas removal of high degree nodes in a large real network can result in major disruption. The usual length of these paths will increase if nodes are removed from a network, resulting in disconnected pairs of nodes and making it more difficult for network agents to communicate.

*Node-level Properties:* The node degree is the number of edges adjacent to that node ($deg_i$). In-degree is the number of edges pointing in to the node ($in\_deg_i$) and out-degree is the number of edges pointing out of the node ($out\_deg_i$). Average neighbor degree refers average degree of the neighborhood ($z_{n,i}$) of each node $i$ is:

$$z_{n,i} = \frac{1}{|N_i|} \sum_{j \in N_i} z_j \quad \ldots\ldots\ldots\ldots\ldots\ldots\ldots\ldots (5)$$

where, $N(i)$ are the neighbors of node $i$ ; $z_j$ is the degree of node $j$ that belongs to $N_i$. In case of weighted graphs, weighted degree of each node can be used [9]. In case of an unweighted graph, the clustering coefficient ($cc_i$) of a node $i$ refers to the fraction of possible triangles that exist through that node:

$$cc_i = \frac{2 T_i}{deg_i * [deg_i - 1]} \quad \ldots\ldots\ldots\ldots\ldots\ldots\ldots\ldots\ldots (6)$$

where, $T_i$ is the number of triangles that exist through node $i$ and $deg_i$ is the degree of node $i$. In case of weighted graphs, this clustering coefficient can be defined as the geometric average of the sub-graph edge weights [10]. The eccentricity of node $i$ is the maximum distance from node $i$ to every other nodes in the graph $G$ ($ecc_i$).

Out of a number centrality measures, betweenness centrality ($BC_i$) of node $i$ is the sum of the fraction of all-pairs of shortest path that pass through node $i$:

$$BC_i = \sum_{x,y \in V} \frac{\theta_{(x,y|j)}}{\theta_{(x,y)}} \quad \ldots\ldots\ldots\ldots\ldots\ldots\ldots\ldots\ldots\ldots (7)$$

where, $V$ is the set of nodes in $G$, $\theta_{(x,y)}$ is the number of shortest $(x, y)$ paths, and $\theta_{(x,y \mid j)}$ is the number of paths that pass through some node $j$ other than $(x, y)$. Please refer to [11-13] for more details. The closeness centrality ($CC_i$) of node $i$ is the reciprocal of the sum of the shortest path distances from node $i$ to all $(n - 1)$ other nodes in the graph $G$:

$$CC_i = \frac{n-1}{\sum_{j=1}^{n-1} \theta_{(j,i)}} \ldots\ldots\ldots\ldots\ldots\ldots\ldots\ldots\ldots\ldots\ldots (8)$$

where, $\theta_{(j,i)}$ is the shortest path distance between node $j$ and node $i$ and $n$ is the number of total nodes in graph $G$. Closeness is normalized by the sum of minimum possible distances of $(n - 1)$ since the sum of the distances depend on the number of nodes in the graph. Higher values of closeness implies higher centrality. Please refer to [14] for details. The eigenvector centrality ($EC_i$) computes the centrality for a node $i$ based on the centrality of its neighbors. The eigenvector centrality for node $i$ is:

$$A\,x = \lambda\,x \ldots\ldots\ldots\ldots\ldots\ldots\ldots\ldots\ldots\ldots\ldots\ldots\ldots (8)$$

where $A$ is the adjacency matrix of the graph $G$ with eigenvalue $\lambda$. Perron–Frobenius theorem suggests that there is a unique and positive solution if $\lambda$ is the largest eigenvalue associated with the eigenvector of the adjacency matrix $A$ [15,16]. Finally, degree centrality for a node is just the fraction of nodes it is connected to. *Other network properties*: Some other common properties are observed in many real networks such as mixing patterns (selective linking), network homophily or similarity, degree correlations, preferential attachment, community structure, network navigation, size of giant components among others [2].

## Subgraph Definition

We followed the following definition [17] to construct the subgraphs:

*A graph $G$ is an ordered triple $(V(G), E(G), \psi_G)$ consisting of a non-empty set $V(G)$ of nodes, a set $E(G)$ of links being disjoint from $V(G)$ and an incidence function $\psi_G$ that associates with each edge of $G$ an unordered pair of (not necessarily distinct) vertices of G. If $\boldsymbol{e}$ is an edge and $\boldsymbol{u}$ and $\boldsymbol{v}$ are vertices such that $\psi_G(e) = uv$, then $\boldsymbol{e}$ is said to join $\boldsymbol{u}$ and $\boldsymbol{v}$; the nodes $\boldsymbol{u}$ and $\boldsymbol{v}$ are called the ends of $\boldsymbol{e}$. A graph $H$ is a subgraph of $G$ $(H \subseteq G)$ if $V(H) \subseteq V(G)$, $E(H) \subseteq E(G)$, and $\psi_H$ is the restriction of $\psi_G$ to $E(H)$. When $H \subseteq G$ but $H \neq G$, we write $H \subset G$ and call $H$ a proper subgraph of $G$. If $H$ is a subgraph of $G$, $G$ is a supergraph of $H$.*

**Supplementary Figure S1 | Snapshots of active subgraphs with their elements**

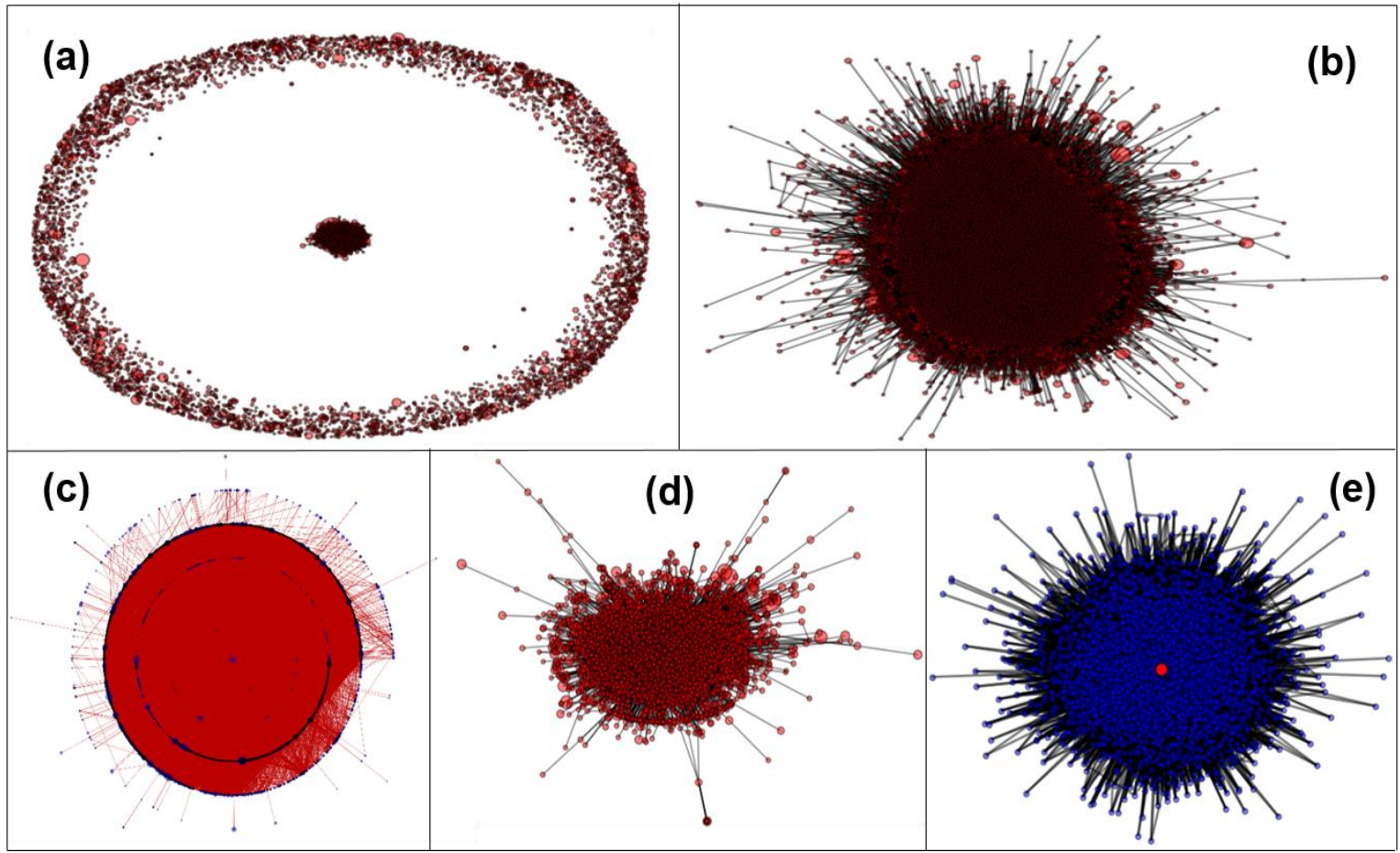

(a) Full subgraph (~ 0.16 M nodes, ~ 14.50 M links, AF ≥ 10, ~ 3.92 M tweets); (b) Largest Connected Component of the subgraph (~ 0.15 M nodes, ~ 11.38 M links, AF ≥ 10); (c) Circular tree visualization of Largest Connected Component (~ 12 K nodes, ~ 0.50 M links, AF ≥ 50) (d) Regular visualization of Largest Connected Component (~ 12 K nodes, ~ 0.50 M links, AF ≥ 50); (e) Largest Hub (AF ≥ 50) ***Node size is proportional to node activity in each case.

## Supplementary Figure S2 | Node-level Properties: Degree and Avg. Neighbor Degree

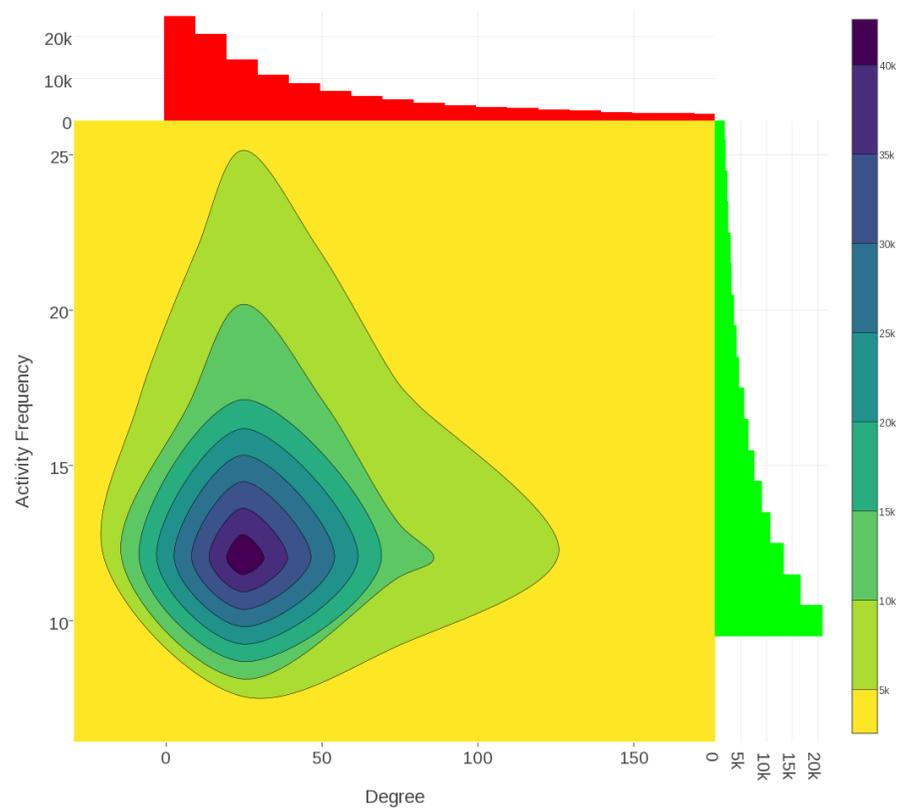

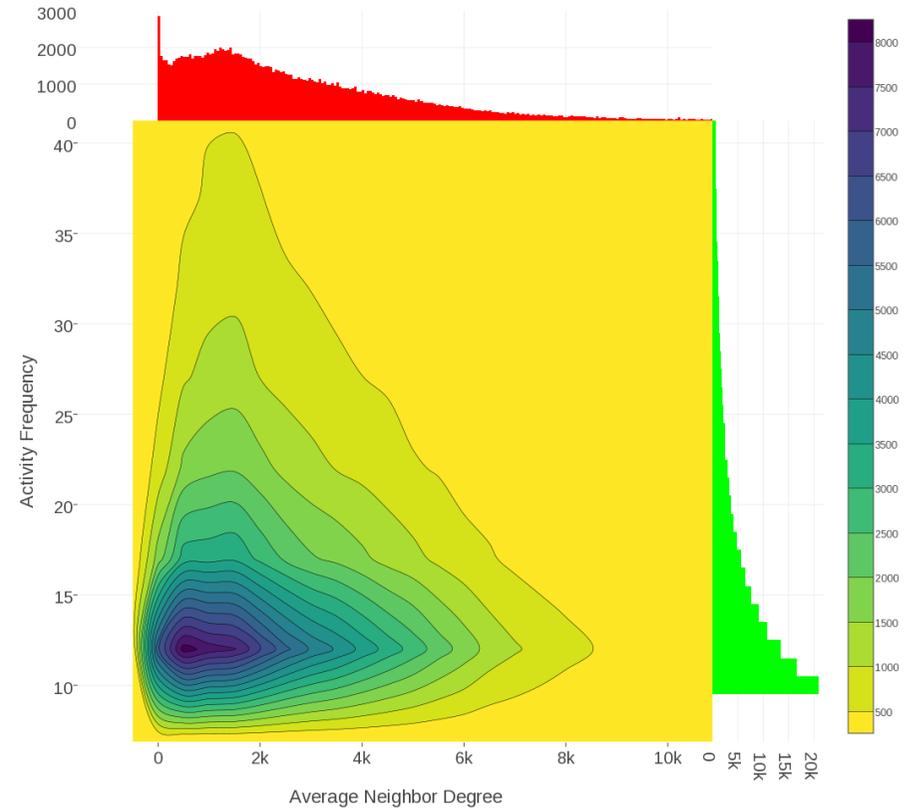

# Supplementary Figure S3 | Node-level Properties: In-degree and Out-degree

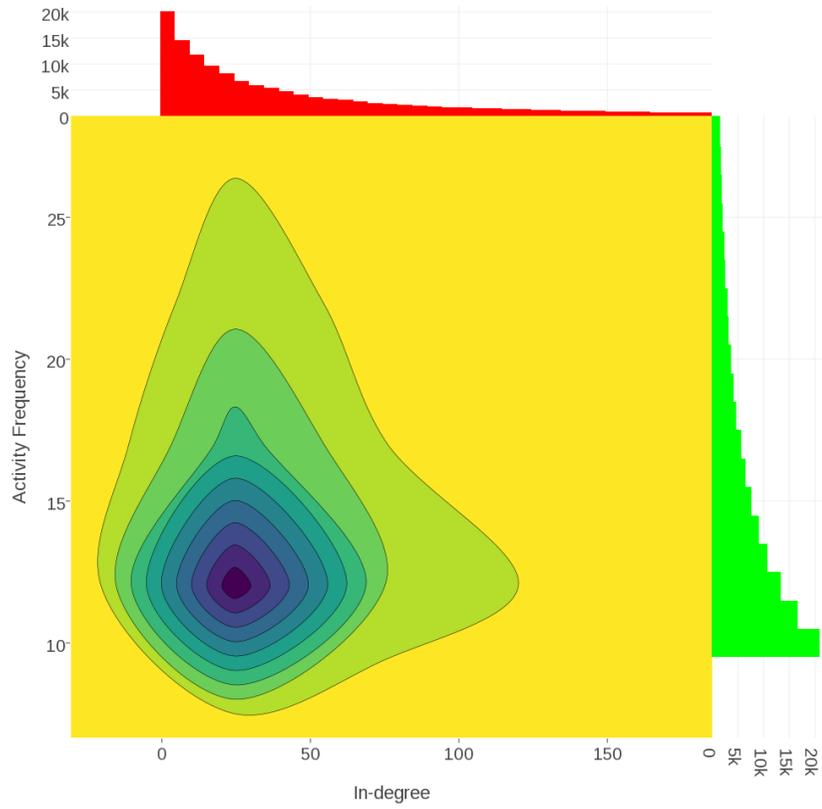

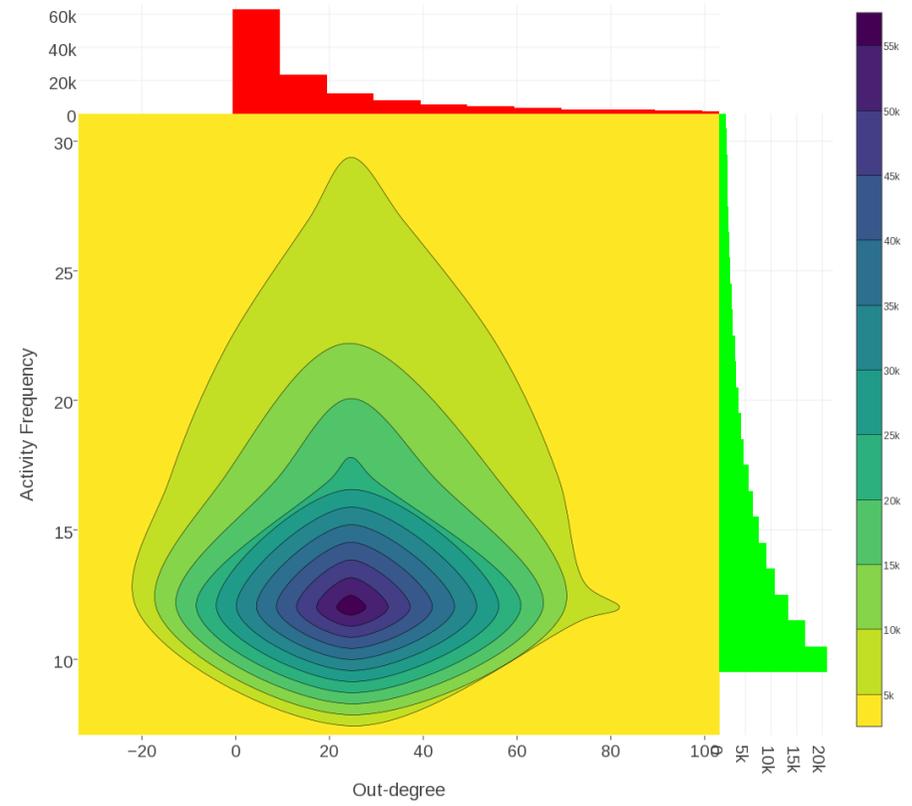

**Supplementary Figure S4 | Node-level Properties: Clustering Coefficent and Eccentricity**

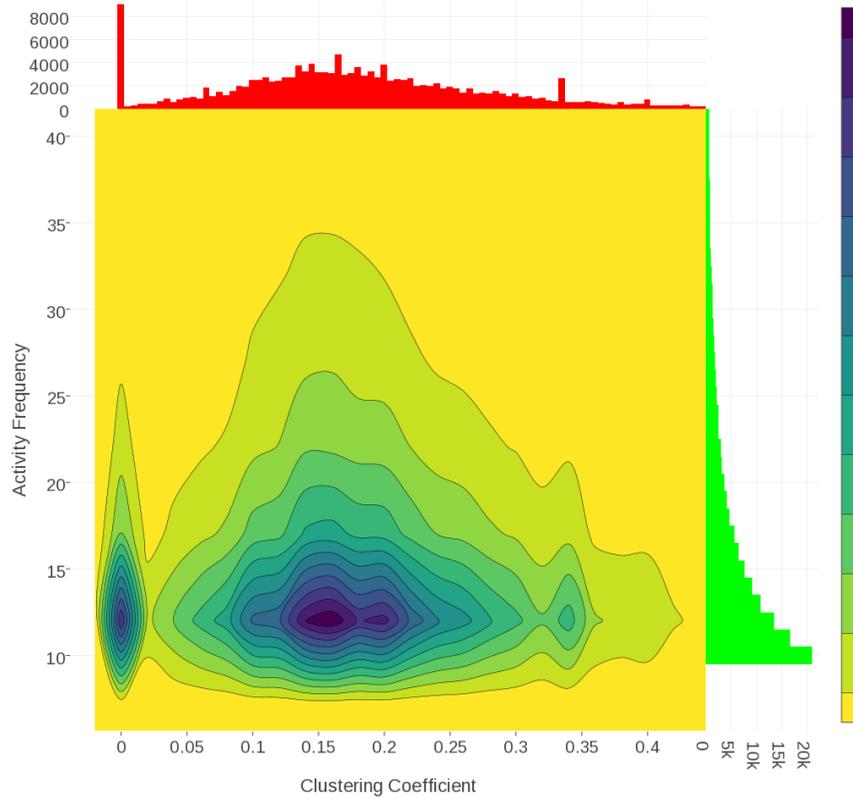

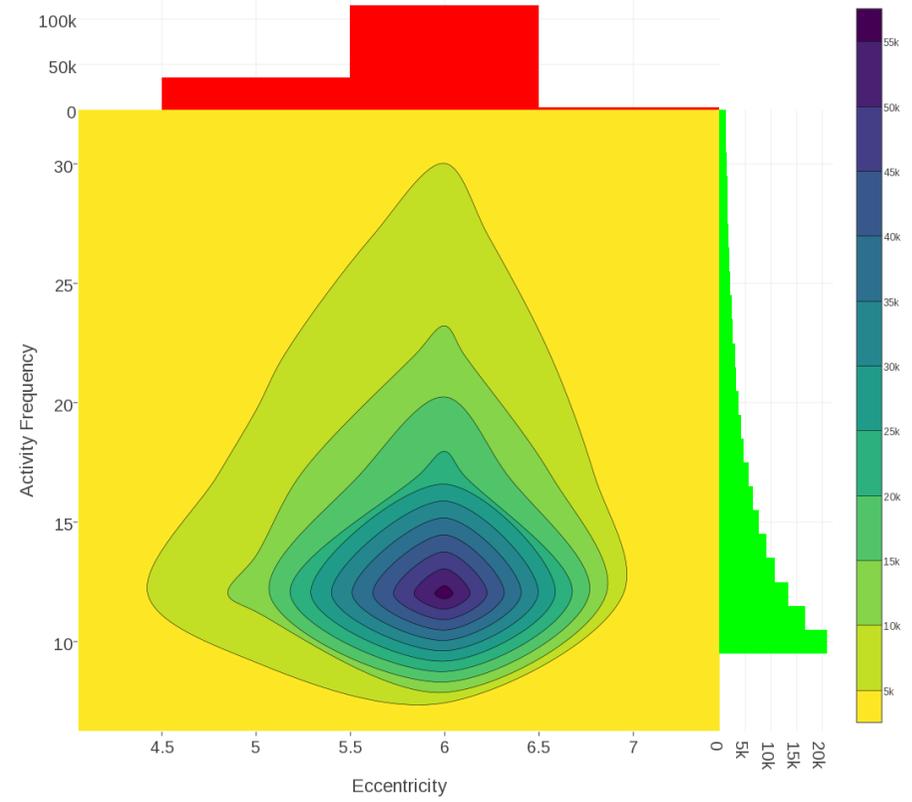

# Supplementary Figure S5 | Node-level Properties: Betweenness Centrality and Closeness Centrality

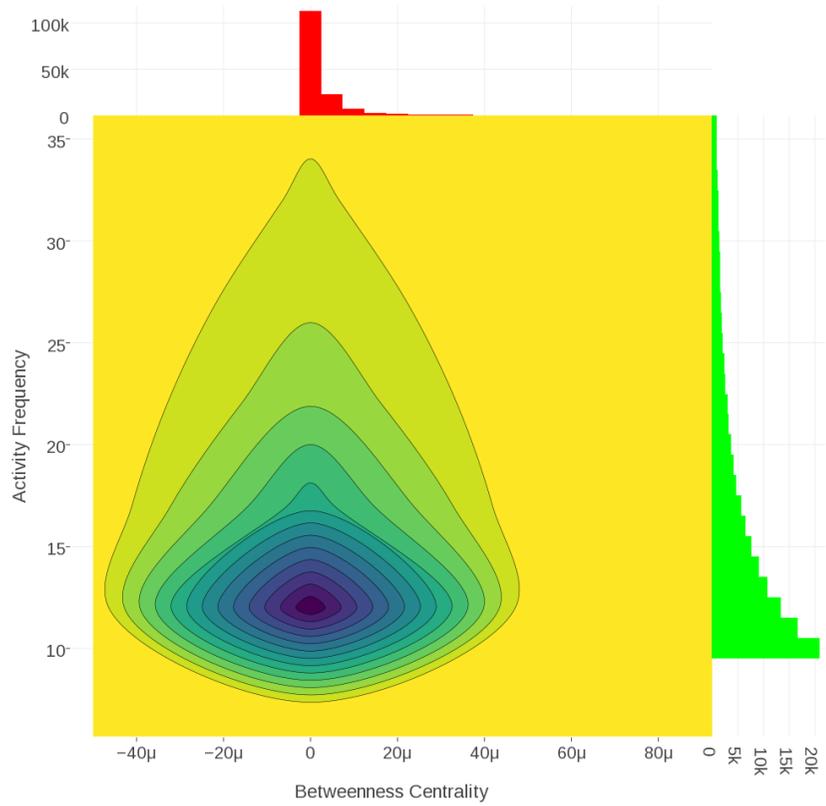

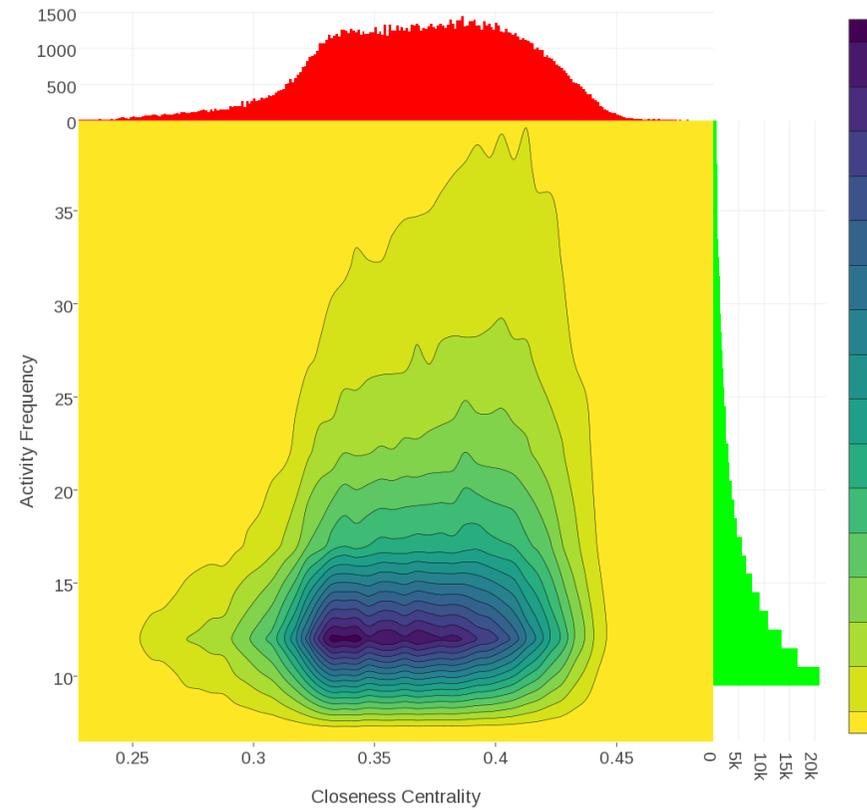

**Supplementary Figure S6 | Node-level Properties: Eigenvector Centrality and Degree Centrality**

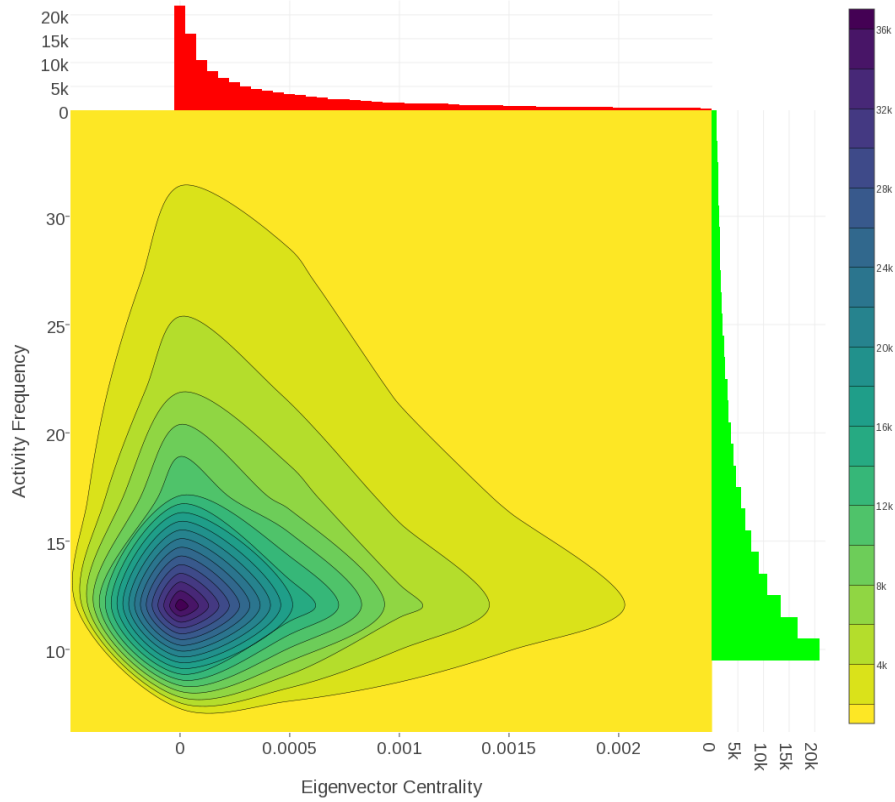

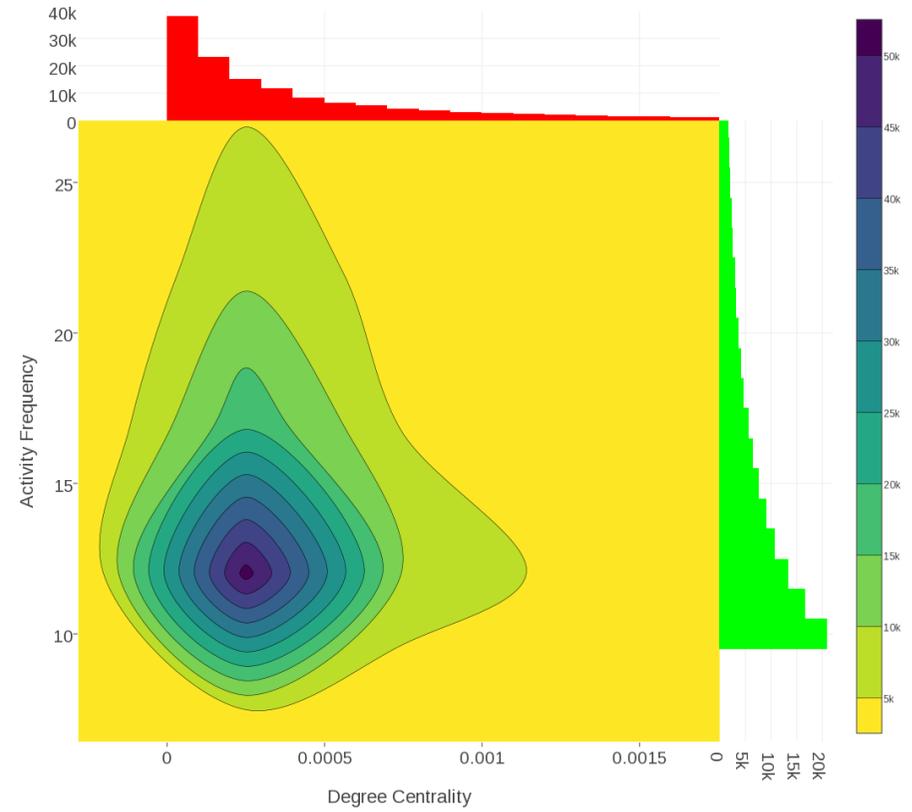

**Supplementary Table S1 | Specific events along the path of Hurricane Sandy**

| Date | Time | Nearby Location | Event |
|---|---|---|---|
| October 22, 2012 | 12:00 UTC | Kingston, Jamaica | Sandy formed and officially assigned name |
| October 24, 2012 | 19:00 UTC | Jamaica | First landfall as a Category 1 hurricane |
| October 25, 2012 | 05:30 UTC | Cuba | Second landfall as a Category 3 hurricane |
| October 29, 2012 | 12:00 UTC | Atlantic City | Re-intensified to the maximum wind speeds |
| October 29, 2012 | 23:30 UTC | Near Brigantine in New Jersey | Final landfall as a post-tropical storm |